\documentclass[aps,pre,showpacs,amsmath,amssymb,amsfonts,superscriptaddress,onecolumn,longbibliography]{revtex4-1}

\usepackage{graphicx}
\usepackage{subfigure}
\usepackage{verbatim}
\usepackage{dcolumn}% Align table columns on decimal point
\usepackage{bm}% bold math
\usepackage{epsf}
\usepackage{color}
\usepackage[colorlinks=true,citecolor=blue,linkcolor=blue]{hyperref}
\usepackage{esint}
\usepackage{epstopdf}

\newcommand{\bla}{bla\\bla\\bla\\bla\\bla}

\definecolor{amethyst}{rgb}{0.6, 0.4, 0.8}
\definecolor{asparagus}{rgb}{0.53, 0.66, 0.42}
\definecolor{fuzzywuzzy}{rgb}{0.8, 0.4, 0.4}

\begin{document}

\title{The validity of the no-pumping theorem in systems with finite-range interactions between particles}

\author{Saar Rahav}
\email{rahavs@technion.ac.il}
\affiliation{Schulich Faculty of Chemistry, Technion-Israel Institute of Technology, Haifa 32000, Israel}

\date{\today}

\begin{abstract}
The no-pumping theorem states that seemingly natural driving cycles of
stochastic machines fail to generate directed motion. Initially derived for single particle systems, the no-pumping theorem
was recently extended to many-particle systems with zero-range interactions. Interestingly, it is known
that the theorem is violated by systems with exclusion interactions.
These two paradigmatic interactions differ by two qualitative aspects: the range of interactions, and
the dependence of branching fractions on the state of the system.
In this work two different models are studied in order to identify the
qualitative property of the interaction that leads to breakdown of no-pumping. A model with finite-range interaction is
shown analytically to satisfy no-pumping. In contrast, a model in which the interaction affects the probabilities of
reaching different sites, given that a particle is making a transition, is shown numerically to violate the no-pumping theorem.
The results suggest that systems with interactions that lead to state-dependent branching fractions do not satisfy the
no-pumping theorem.
\end{abstract}

 \pacs{05.60.-k, 05.70.Ln, 82.20.-w}

\maketitle

\section{Introduction}
\label{sec:intro}

One of the prominent technological trends of the last century is miniaturization. The race to build
 smaller devices and machines has led to the realization that when a system is sufficiently small
its properties change qualitatively. Molecular machines are one such fascinating example of the notion that ``small is different'' \citep{howard_mechanics_2001-1,coskun_great_2011,kay_synthetic_2007,kottas_artificial_2005,michl_molecular_2009,feringa_art_2007}.
Macroscopic machines work in an ordered and predictable manner, since their interaction with their environment
is relatively weak and does not disrupt their operation. In contrast, microscopic machines are strongly agitated by their surroundings  \cite{Bustamante2005},
resulting in large fluctuations in their dynamics.

The coupling of a small system to a thermal environment is naturally described in terms of
stochastic equations of motion, expressing our inability to predict the dynamics of the environment.
Our theoretical understanding of such systems significantly improved in the last two
decades, following the discovery of the celebrated fluctuation theorems \cite{Evans1993,Evans1994,Gallavotti1995,Kurchan1998,Lebowitz1999,
Seifert2005, Rahav2013,Harbola2014,Jarzynski1997,Crooks1998}. A theory, termed stochastic
thermodynamics, allows to assign thermodynamic interpretation to single realizations of a
nonequilibrium process \cite{Sekimoto1997,Seifert2012}. It can be viewed as extending thermodynamics
to small out-of-equilibrium systems.

The small size of molecular machines means that
finding reliable ways of driving them is not a trivial task.
Molecular machines of biological origin typically exhibit coupling
between mechanical motion and a chemical reaction. Many biological motor proteins are driven by the chemical
potential difference between ATP and its hydrolysation products.
This chemical potential
difference changes on a time scale which is much larger than the time it takes
such a motor to make a step or complete a cycle. Such machines are therefore
driven by an effectively time-independent chemical driving force.

Not all microscopic machines are of biological origin. Many research groups are designing and building
artificial nano-machines in their labs \citep{kay_synthetic_2007,kottas_artificial_2005,michl_molecular_2009,feringa_art_2007}.
These artificially designed machines can be driven by mechanisms which are very different from those of their biological counterparts. One interesting
driving mechanism involves a cyclic variation of external parameters, which then results in some kind of directed motion. Thermal ratchets are one well known
type of a periodically driven system where diffusion plays an integral part of the dynamics \cite{Reimann2002}.
But cyclic variation of parameters can also be used to drive systems modeled
by a set of discrete states, where diffusion is not as prominent. The latter class of systems is often
referred to as stochastic pumps \citep{sinitsyn_universal_2007,sinitsyn_stochastic_2009,chernyak_quantization_2012,parrondo_reversible_1998,astumian_adiabatic_2007,sokolov_perturbation_1999,astumian_adiabatic_2003,rahav_extracting_2011,sinitsyn_stochastic_2009,astumian_stochastic_2011} since they are driven in a way which is similar to everyday, macroscopic pumps.

In a beautiful experiment Leigh {\it et al.} have built an
artificial molecular machine that operates as a stochastic pump \cite{leigh_unidirectional_2003}. A complex of interlocked ring-like molecules was synthesized.
Relative motion of the ring-like molecules in this machine was achieved
by a cycle of local chemical changes at binding sites along a larger molecule. These can be modeled
by a time variation of local binding energies, so that the smaller ring-like molecules
makes thermally activated transitions in a time-dependent energy landscape. Surprisingly, it turns out
that temporal variation of local binding energies does not result in directed motion, even when the time-ordering of
the variation breaks the symmetry between directions \cite{rahav_directed_2008,chernyak_pumping_2008}.
One needs to also vary local barriers during the cycle to generate directed motion.
This result is known as the no-pumping theorem. It's non intuitive nature and simplicity have generated
some interest  \citep{horowitz_exact_2009,mandal_hybrid_2012,maes_general_2010,mandal_proof_2011,ren_duality_2011}.

In many-particle stochastic pumps interactions between the particles may have considerable
effect on the resulting dynamics. In the context of no-pumping one may wonder whether interactions may result
in violation of the theorem. Already in the experiment of Leigh {\it et al.} it was noted
that exclusion interaction allows one to achieve directed motion in a pump driven by
variation of site energies alone \cite{leigh_unidirectional_2003}, meaning that
exclusion interaction breaks the no-pumping theorem.
(See also \cite{astumian_adiabatic_2003} for a theoretical discussion.) In contrast,
several recent papers studied many-particle pumps with local, or zero-range, interactions and
found that the no-pumping theorem holds \cite{asban_no-pumping_2014,mandal_unification_2014,Asban2015}.

It is not surprising that the two types of interactions used so far in studies of stochastic pumps
are exclusion and zero-range. These two interactions serve as paradigmatic models of many-particle
nonequilibrium steady states, and have been extensively investigated in that context \cite{spohn_large_1991,kipnis_hydrodynamics_1989,evans_nonequilibrium_2005,levine_zero-range_2005,kipnis_scaling_1999}.
But the current state of knowledge, where it is known that exclusion interaction violates no-pumping, while zero-range
interaction do not, leaves an open question. Which property of particle-particle interactions leads
to the breakdown of the no-pumping theorem? A closer inspection of exclusion and zero-range interactions reveals that
they differ by two qualitative aspects.
One is the range of interaction. Exclusion interaction is a nearest neighbor interaction, and therefore has a finite range, while zero-range is a local
interaction. The other qualitative aspect has to do with the conditional probability of a particle to reach different sites
when it is known that this particle is making a transition. The so-called branching fractions expressing this
conditional probability are time-independent in the zero-range pump, while they gain time dependence on average -
through their dependence on the location of neighboring particles - for exclusion interactions. Which of these
qualitative aspects of interaction is responsible for the violation of the non-pumping theorem?

In this paper I answer this question by studying the validity of no-pumping in models with finite range
interactions that still exhibit similarities to zero-range interactions. I find that if the interaction is such
that the branching ratios are fixed, for instance because they are independent of the state of the system, the no-pumping theorem still holds.
I also study numerically a two-particle system
with a more complicated interaction which results in branching fractions that depend on the configuration of particles, and find that the no-pumping theorem breaks down.

The structure of the paper is as follows. In Sec. \ref{sec:anal} a model of a stochastic pump with a particular form
of finite-range interaction between all particles in the system is constructed. It is then shown that the no-pumping theorem
holds for this model. A model of a two-particle stochastic pump with more general interaction is studied numerically
in Sec. \ref{sec:num}, and is found to violate the theorem. In Sec. \ref{sec:disc} the qualitative differences between the interactions
are discussed, and the aspect of interaction that leads to breakdown of the no-pumping theorem is identified.

\section{No-pumping with finite-range interactions}
\label{sec:anal}

In this section I define a particular model of stochastic pumps with non-local interactions. I then demonstrate that this
  model satisfies the no-pumping theorem. The model, which is described below,
can be viewed as a generalization of the zero-range interaction studied
in Refs. \cite{asban_no-pumping_2014,mandal_unification_2014,Asban2015}. It also
have similarities with models that were studied in the context of pair factorized steady states \cite{Evans2006,Waclaw2009}.

Consider a system of $N$ particles and $M$ sites. The system's
state can be specified by denoting the precise location of each particle, or alternatively
by considering the occupation of each site. In this section I will adopt the latter notation since I will be interested
in the total particle current, and therefore there is no need to distinguish between different particles. The states of the system are
denoted by the occupation numbers
$ \bar{n}\equiv \left( n_1,n_2, \cdots , n_M\right)$, with the constraint $\sum_{i=1}^M n_i=N$. When a particle resides
in site $i$ it has a local (or site) energy $E_i (t)$, which can be controlled externally. In addition, all the particles interact
with each other, via some finite-range many-body interaction $U(\bar{n})$. Crucially, this interaction depends on the locations of all the
particles in the system. As a result every state
of the system has an overall energy of
\begin{equation}
E_{\bar{n}}(t)=\sum_{i=1}^M E_i (t) n_i +U(\bar{n}).
\label{eq:defmbenergy}
\end{equation}
The interaction $U(\bar{n})$ can have any finite value, as long as it does not destroy or add metastable states (that is, sites)
to the system.

A model of a stochastic pump used to study no-pumping should relax to equilibrium when not driven. The
equilibrium distribution is the Boltzmann distribution, which is given by
\begin{equation}
\label{eq:boltzmann}
  P_{\bar{n}}^{eq} =\frac{N!}{n_1! n_2! \cdots n_M!} \frac{e^{-\beta E_{\bar{n}}}}{\cal Z},
\end{equation}
where ${\cal Z} \equiv \sum_{\bar{n}|_{\sum_i n_i = N}}\frac{N!}{n_1! n_2! \cdots n_M!} e^{-\beta E_{\bar{n}}}$ is the corresponding canonical partition function.
The rates of transitions between many-particle states are chosen so that this Boltzmann distribution
is indeed the equilibrium distribution. To do so one defines operators $\hat{b}_i^\pm$ that
add or subtract a particle to site $i$. A transition of a particle from site
$i$ to site $j$ then connects the states $\bar{n}_1$ and $\bar{n}_2=\hat{b}_j^+ \hat{b}_i^- \bar{n}_1$.
While making this transition the system must overcome an energetic barrier, which has the form
\begin{equation}
\label{eq:defmbbarrier}
W_{\bar{n}_2, \bar{n}_1} \equiv E_{\hat{b}_i^- \bar{n}_1} +B_{j,i}= \sum_{j=1}^M E_j  \left[n_{1j}-\delta_{ij} \right]+U(\hat{b}_i^- \bar{n}_1) +B_{j,i},
\end{equation}
where $B_{i,j}=B_{j,i}$ are local energy barriers for transitions between sites.
(When $n_{1i}$=0 this expression is not well defined, but this is not a problem because the transition rate is defined later in a way that
ensures that it vanishes for transitions out of an empty site.)
The energetic barrier for transitions is therefore a sum of the energy of the $N-1$
stationary particles and the local barrier for the transition.
The transitions are assumed to be thermally activated, with rates
\begin{equation}
\label{eq:defrates}
R_{\bar{n}_2, \bar{n}_1} = n_{1i} e^{-\beta \left(W_{\bar{n}_2, \bar{n}_1}- E_{\bar{n}_1}\right)}=n_{1i} e^{-\beta \left[B_{j,i}+ U(\hat{b}_i^-\bar{n}_1)-E_i-U(\bar{n}_1) \right]}
\end{equation}
for $\bar{n}_2=\hat{b}_j^+ \hat{b}_i^- \bar{n}_1$, while $R_{\bar{n}^\prime, \bar{n}}=0$ when the states $\bar{n}^\prime, \bar{n}$ are not
connected by a single transition.
The factor of $n_{1i}$ expresses the fact that each of the particles in site $i$ can make the transition, and
they are all equally likely to do so. Crucially, these many-particle barriers are symmetric, namely $W_{\bar{n}_2, \bar{n}_1}=W_{\bar{n}_1, \bar{n}_2}$.
The system's probability distribution evolves according to a master equation
\begin{equation}\label{eq:master}
  \frac{d P_{\bar{n}}}{dt}=\sum_{{\bar{n}}^\prime} R_{\bar{n}, \bar{n}^\prime}  P_{\bar{n}^\prime},
\end{equation}
where $R_{\bar{n}, \bar{n}} \equiv - \sum_{{\bar{n}}^\prime \ne \bar{n}} R_{\bar{n}^\prime, \bar{n}}$ ensures conservation of probability.

The transition rates (\ref{eq:defrates}) are consistent with the Boltzmann distribution (\ref{eq:boltzmann}) as they satisfy
the detailed balance condition $R_{\bar{n}^\prime, \bar{n}}P_{\bar{n}}^{eq}=R_{\bar{n}, \bar{n}^\prime}P_{\bar{n}^\prime}^{eq}$. This is true for
any symmetric choice of barriers. The specific choice made in Eq. (\ref{eq:defmbbarrier}) neglects the interaction between
the particle that makes a transition and other particles. It is this property of the many-particle barriers which makes the model qualitatively similar to the zero-range process.
For this representation
of many-particle states the total particle current between sites $i$ and $j$
is expressed by \cite{mandal_unification_2014}
\begin{equation}
\label{eq:totcurrent}
  J_{j,i} (t)= \sum_{\bar{n}_1} J_{\bar{n}_2,\bar{n}_1}(t)=\sum_{\bar{n}_1} \left[ R_{ \bar{n}_2,\bar{n}_1} P_{\bar{n}_1}(t)- R_{\bar{n}_1,\bar{n}_2}   P_{\bar{n}_2}(t) \right],
\end{equation}
where in all the terms in the sum $\bar{n}_2=\hat{b}_j^+ \hat{b}_i^- \bar{n}_1$. It should be noted that $J_{\bar{n}_2,\bar{n}_1}(t)=R_{ \bar{n}_2,\bar{n}_1}(t) P_{\bar{n}_1}(t)- R_{\bar{n}_1,\bar{n}_2} (t)  P_{\bar{n}_1}(t)$ is the flux of probability between the configurations $\bar{n}_2,\bar{n}_1$, whereas $J_{j,i}(t)$ is the total
particle current between sites $i$ and $j$.

This model of a jump process relaxes to equilibrium when the site energies and barriers are not varied in time. It can be operated as a pump
by varying the site energies $E_{i}(t)$ and local barriers $B_{i,j} (t)$ periodically in time.
 The Floquet theorem ensures that in the long time limit the system converges to a periodic
state, $P^{ps}_{\bar{n}}(t+\tau) = P^{ps}_{\bar{n}}(t)$ \cite{Talkner1999}. In this periodic state particles
may preferentially move in different directions at different times. Directed
motion is achieved when more transitions are accumulated in one direction than in its opposite. The time-integrated currents
\begin{equation}
\label{eq:defintc}
  \Phi_{j,i}^{ps} \equiv \int_{t}^{t+\tau} dt^\prime J_{j,i}^{ps} (t^\prime)
\end{equation}
can be used as a measure for the directed motion of the particle in the pump. When $ \Phi_{j,i} \ne0$ for at least one pair of sites $(i,j)$ directed motion is achieved
and the system can be operated as a thermodynamic engine, and perhaps do some useful work against a resisting force \cite{rahav_extracting_2011}. (In the following
I suppress the superscript $ps$, as I will {\em only} consider systems that are in this periodic state.)

The no-pumping theorem states that both the site energies, $E_i (t)$, and the local barriers $B_{i,j} (t)$, must be varied in time
to generate directed motion.
Variation of only site energies, or alternatively of only local barriers, will not result in currents that have a preferred direction (after a full cycle).
The no-pumping theorem is non-intuitive because variation of the site energies can be done
in a way that gives the system a preferred direction. As an example consider a system with three connected sites, $A,B$ and $C$.
The site energies can be varied so that during the cycle the most binding site changes from $A$ to $B$, then to $C$, and finally back to $A$.
Such a variation gives a preferred direction to the driving protocol, but due to the no-pumping theorem we know that the particles
will not respond in a way that exhibit preferred direction after a full cycle.
I now demonstrate that this theorem, which was shown to be valid for non-interacting particles and for zero-range interactions, is also valid for
the model described above. Crucially, this model includes an interaction $U(\bar{n})$ which clearly has a finite range, as it allows for interactions
between particles in all sites.

The derivation of the no-pumping theorem for the case of time independent local energies is straightforward. The system relaxes to
the time-independent equilibrium distribution given by Eq. (\ref{eq:boltzmann}). This distribution is undisturbed by the time dependence of
the local barriers. Once the system has relaxed to this distribution all the currents vanish, $J_{i,j} (t)=0$. There is no directed motion.

The derivation for time-independent local barriers is based on
the approach developed by Mandal and Jarzynski for single-particle
stochastic pumps \cite{mandal_proof_2011}. This approach was later extended to many-particle stochastic pumps
with zero range interactions \cite{asban_no-pumping_2014,mandal_unification_2014}. Below I show that it applies
to the model of Eqs. (\ref{eq:boltzmann})-(\ref{eq:defrates}) with very little modifications.
This proof of the
no-pumping theorem is based on a derivation of two sets of equations, termed conservation laws and cycle equations. These two sets
are then shown to be incompatible with each other, so that the only possible solution is one without
preferred direction for particle flow ($ \Phi_{j,i} =0$ for all transitions). The introduction of the finite-range interaction only modifies the derivation
of the cycle equations. To avoid repeating already published arguments I will focus this part of the derivation
and will only heuristically describe the rest of the steps.

The conservation laws follow from the periodicity of the driving protocol and the fact that the number of particles in the system
is conserved. In fact, their derivation does not depend
on the form of the interaction, as they just express the fact that particles do not appear or disappear. They have been
derived before for many-particle pumps with zero-range interaction, and that derivation remains be valid for the model studied here. An example of the derivation
which uses site occupations to describe states of a system can be found in Ref. \cite{mandal_unification_2014}. The
 conservation laws are given by
\begin{equation}
\label{eq:claws}
  \sum_{j\ne i} \Phi_{j,i}=0.
\end{equation}
Equation (\ref{eq:claws}) simply expresses the fact that the total flux of particles out of site $i$ must vanish.
It should be noted that Ref. \cite{mandal_unification_2014} considered an open system in which particles could enter or leave to external particle
reservoirs, and as a result also had somewhat different
conservation laws for sites connected to particle reservoirs. There are no such sites here.
By themselves the conservation laws (\ref{eq:claws}) do not mean that all integrated currents must vanish. Instead they imply that particles must flow in closed cycles of transitions. Any other flow pattern will result in accumulation of particles and will
not be consistent with the time-periodicity of the dynamics. This is the periodically-driven analogue of the well known network theory of steady-states \cite{Schnakenberg1976}.

The other set of equations is called cycle equations since they refer to particle currents along
a closed cycle of transitions between sites. As pointed out by Mandal and Jarzsynki, these equations essentially result
from the detailed balance form of the transition rates.
The derivation of cycle equations is slightly modified by the presence of the finite-range interaction. Consider the combination
\begin{equation*}
  e^{\beta B_{j,i}} J_{\bar{n}_2 ,\bar{n}_1}(t) = n_{1i} e^{-\beta \left[-E_i(t) - U(\bar{n}_1)+U(\hat{b}_i^- \bar{n}_1) \right]} P_{\bar{n}_1} (t)-n_{2j} e^{-\beta \left[ -E_j(t)-U(\bar{n}_2 )+U(\hat{b}_i^- \bar{n}_1)\right]} P_{\bar{n}_2}(t) ,
\end{equation*}
where we again used the notation $\bar{n}_2 = \hat{b}_j^+ \hat{b}_i^- \bar{n}_1$, so that $n_{2j}=n_{1j}+1$.
Cycle equations will emerge when the summation of such terms from transitions between neighboring sites will show telescopic cancelations. For instance, if one considers also
the $j\rightarrow k$ transition between many particle states $\bar{n}_2$ and $ \bar{n}_3 =\hat{b}_k^+ \hat{b}_j^- \bar{n}_2 = \hat{b}_k^+ \hat{b}_i^- \bar{n}_1$ one finds
\begin{equation*}
  e^{\beta B_{k,j}} J_{\bar{n}_3,\bar{n}_2}(t) = n_{2j} e^{-\beta \left[ -E_j(t)-U(\bar{n}_2)+U(\hat{b}_i^- \bar{n}_1)\right]} P_{\bar{n}_2}(t)-n_{3k} e^{-\beta \left[ -E_k(t)-U(\bar{n}_3)+U(\hat{b}_i^- \bar{n}_1)\right]} P_{\bar{n}_2}(t).
\end{equation*}
Note that in both transitions the configuration of the stationary particles is $\hat{b}_i^- \bar{n}_1$, leading to the appearance of the same factor of $U(\hat{b}_i^- \bar{n}_1)$
in all the terms.

When the two contributions above are added the negative term in the first will cancel the positive one in the second. One concludes that a complete cancellation
of terms
will happen for any closed cycle $i_1\rightarrow i_2 \rightarrow \cdots i_l \rightarrow i_1$ whose many-particle transitions connect a series of many-particle states $\bar{n}_1, \bar{n}_2, \cdots ,\bar{n}_l$, resulting in
\begin{equation}\label{eq:intermediate}
  e^{\beta B_{i_2, i_1}}J_{\bar{n}_2,\bar{n}_1}(t)+e^{\beta B_{i_3, i_2}}J_{\bar{n}_3,\bar{n}_2}(t)+\cdots = 0.
\end{equation}
One can then sum over all many particles states $\bar{n}_1$, use Eq. (\ref{eq:totcurrent}) to recast the result
in terms of total particle currents between sites,
and then integrate over time.
This series of steps results in the so-called  cycle equations for the net flux of particles between sites
\begin{equation}
\label{eq:cycles}
  e^{\beta B_{i_2,i_1}} \Phi_{i_2,i_1} +  e^{\beta B_{i_3,i_2}} \Phi_{i_3,i_2}+\cdots =0.
\end{equation}
Crucially, once such cycle equation holds for every closed cycle of transitions between sites.

Note that Eq. (\ref{eq:cycles}) is valid only for time-independent barriers. When the barriers $B_{i,j} (t)$ are time dependent integration of Eq. (\ref{eq:intermediate}) over time can no longer be expressed in terms of the integrated fluxes, $\Phi_{i,j}$. It is also crucial to notice that this derivation
relies on the fact that the same prefactor of $e^{\beta B_{i,j}}$ appears for all currents in cycles that correspond to a transition $i\rightarrow j$
independently of the many-body state of the system. This allows to sum cycle equations over the possible particle configurations and replace
the probability currents with the total particle currents between sites. The derivation of Eq. (\ref{eq:cycles}) will break down when either one of these system
properties is modified.

The form of the conservation laws (\ref{eq:claws}) and cycle equations (\ref{eq:cycles}) derived here is {\em identical} to those
found for single particle pumps in Ref. \cite{mandal_proof_2011} and for zero-range interactions in Refs. \cite{asban_no-pumping_2014} and \cite{mandal_unification_2014}. As a result the rest of the proof of Mandal and Jarzynski \cite{mandal_proof_2011} is also valid for the
interacting model considered here. I will not repeat that part of the argument here. Heuristically the proof is based on the observation that cycle equations
and conservation laws try to impose different signs on the flux of particles. Conservation laws try to impose the same sign on all fluxes in a cycle. In contrast,
due to the positivity of the $e^{\beta B_{j,i}}$ factors, the cycles equations mandate different signs for some $\Phi$'s in a cycle.
Mandal and Jarzynski then showed that this means that the equations are incompatible and the only solution is one where $\Phi_{j,i}=0$ for all transitions,
even if the system have more than one cycle of transitions.

One sees that the existing derivation of the no-pumping theorem can be extended to include finite-range interactions of the form (\ref{eq:defmbenergy}) as long
as the transition rates are given by Eq. (\ref{eq:defrates}). This means that it is not the range of interaction that leads to the violation
of the no-pumping theorem in the exclusion process. Instead this result suggests that the ability of the interaction to affect the likelihood of different transitions from the
same site  at different times
is the crucial property of the interaction that may lead to directed motion.

\section{Numerical example}
\label{sec:num}

The model studied in Sec. \ref{sec:anal}
can be criticized as being somewhat artificial. The main reason for such criticism is an assumption
that was
made about the interaction between particles. Specifically, it was assumed that particles residing in sites interact with
each other, but not with a particle making a transition. While this assumption seems
 reasonable and internally consistent for zero-range interactions, where particles affect each other only if they are
 in the same site,
 it is somewhat problematic for finite-range interactions, since
 it means that the interaction is felt before and after, but not during the transition.
  In this section
I consider a model of a stochastic pump with interactions between {\em all} particles, including ones in the midst
of a transition.

\begin{figure}
\includegraphics[scale=0.45]{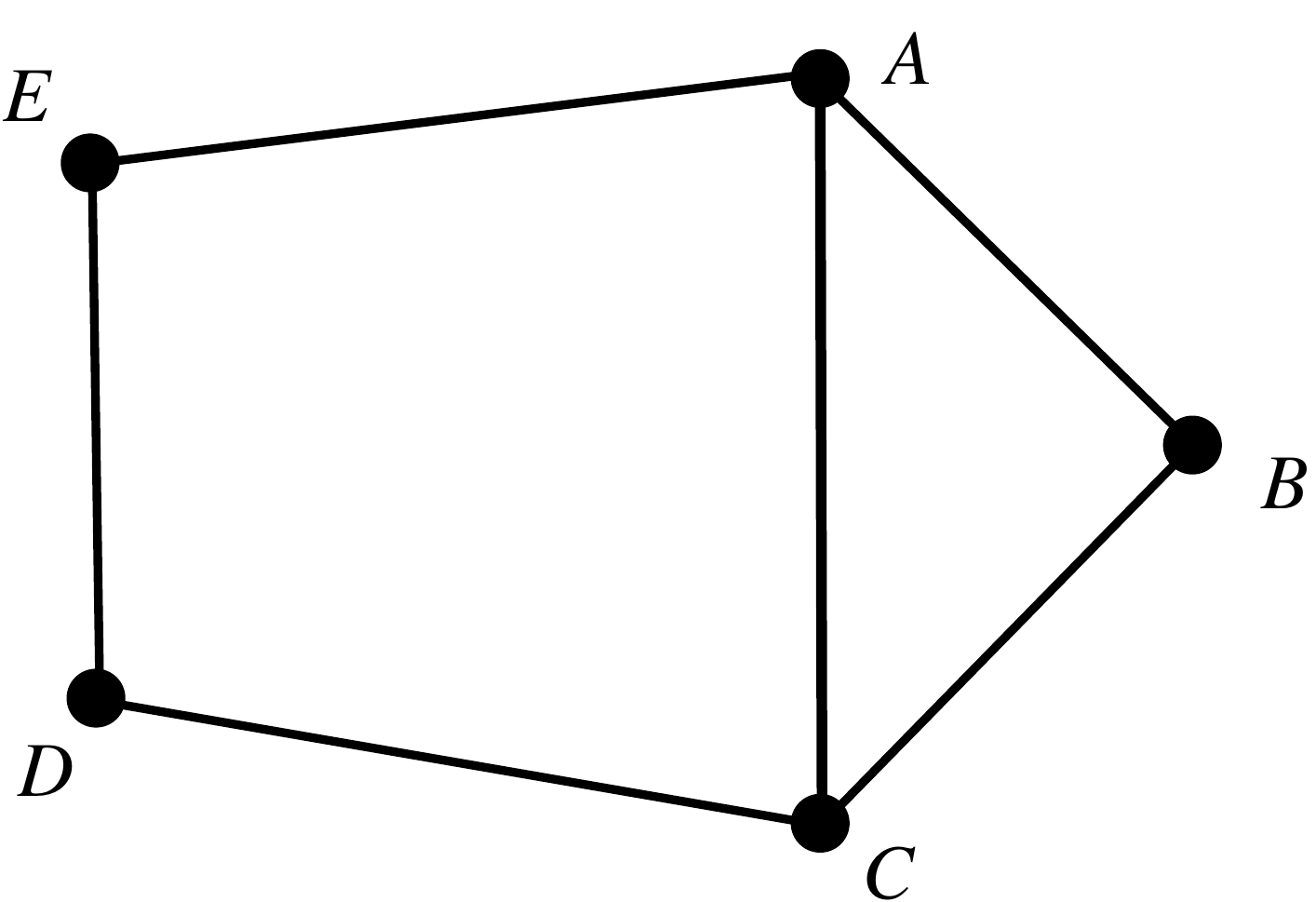}

\caption{Depiction of the five-site system as a graph. The vertices correspond to the different sites, whereas the links correspond to the possible (bidirectional) transitions between sites. \label{fig:graph}}
\end{figure}

The model is chosen to have five sites. These sites, and the topology of transitions between them, are depicted in Fig. \ref{fig:graph}.
The vertices of the graph denote the sites, while the links correspond to the possible transitions between sites. Missing links
are meant to represent forbidden or impossible transitions. This specific five-site configuration was chosen
due to a combination of two reasons: i) It has more than one closed cycle of transitions, and ii) it allows for different ``distance''
between particles, measured as the minimal number of jumps needed to bring one of them to the other. The latter property will be used in the
definition of the interaction, as explained below.

In what follows I will focus on a system of two particles moving in the 5 sites depicted in Fig. \ref{fig:graph}.
When the number of particles is much smaller than the number of sites the easiest way of denoting the
state of the system is by listing the location of each particle. (I assume here that the particles
are distinguishable.)
The model has $25$ many-particle states and a transition matrix of size $25 \times 25$
whose off diagonal entries are the transition rates. Diagonal elements are chosen so that probability is conserved.
Limiting the number of particles in the system to two restricts the type of interactions that one can study.
Nevertheless, I expect that the intuition gained from this model can be applied to more general types of periodically
driven interacting systems. I assume that one can neglect events where both particles make a simultaneous transition.

The interaction between particles is chosen to depend on the ``distance'' between them. This distance is defined as the
smallest number of transitions needed to bring them to the same site. For the graph depicted in Fig. \ref{fig:graph}
this interaction can get one of three values, $U_0$, $U_1$ or $U_2$. For instance $U(CC)=U_0$ while $U(BE)=U_2$.

The proof of no-pumping presented in Sec. \ref{sec:anal} was valid for a finite-range interaction between the particles.
But the interaction used was of a special type.
During a transition only the stationary particles were assumed to feel the interaction. The jumping particle
was unaffected.
As discussed above it makes more sense that a finite-range interaction between particles is also felt
during the transition.
 In this section I therefore consider such an interaction, which is also assumed to depend
on the ``distance'' between particles, and therefore can have the three values $U_{1/2}$, $U_{3/2}$ and $U_{5/2}$. As will be
evident later, this part of the interaction can result in the breakdown of the no-pumping theorem.

These are all the ingredients needed to fully define the model and determine the transition rates between states. Let us take, for
example, the transition from state from state $AB$ to state $CB$. In this transition
the first particle jumps from $A$ to $C$, while the second particle resides in site $B$. The many particle energy
of the initial state is $E_{AB} (t) = E_A (t) + E_B (t)+U_1$, while the many-particle barrier is $ W_{CB,AB} (t)= E_B(t) +B_{A,C}+U_{3/2}$.
The resulting transition rate is given by
\begin{equation}
  R_{CB,AB}(t)=\exp \left[-\beta \left(  W_{CB,AB} (t)-E_{AB}(t)\right) \right]=\exp \left[-\beta \left( U_{3/2}-U_1+B_{A,C}-E_A(t)\right) \right].
\end{equation}
The rest of the transition rates are determined using the same prescription. They are too numerous to be
listed explicitly here.
Note that all the many particle barriers are symmetric, since the same particle configuration appears for a transition and its reverse.
For the example above $W_{CB,AB}=W_{AB,CB}$. This ensures that the system relaxes to equilibrium when parameters are not varied in time.

This two-particle system is operated like a pump by cyclic variation
of the site energies $E_X (t)$ and/or the local barriers $B_{X,Y} (t)$.
The master equation that describes the dynamics is solved numerically, using fourth order Runge-Kutta integrator.
The system is propagated until it settles into a periodic state, and then the time integral of the
total particle currents between sites is calculated. These total particle currents are given by sums of
fluxes of the many-particle transitions in which a particle makes a specific transition. For example the total particle current for the $A \rightarrow B$ transition
is given by
\begin{equation}
\label{eq:totcurrent2}
  J_{A \rightarrow B}^{tot} = \sum_{X=A,B,\cdots E} \left[ J_{A\rightarrow B,X}+J_{X,A\rightarrow B}\right].
\end{equation}
The cyclic variation of parameters results in directed motion when $\Phi_{X,Y} (\tau)= \int_{t}^{t+\tau}dt^\prime J_{Y \rightarrow X}^{tot} (t^\prime) \ne 0$
for any of the $j \rightarrow i$ transitions.

\begin{figure}[h]
\includegraphics[scale=0.45]{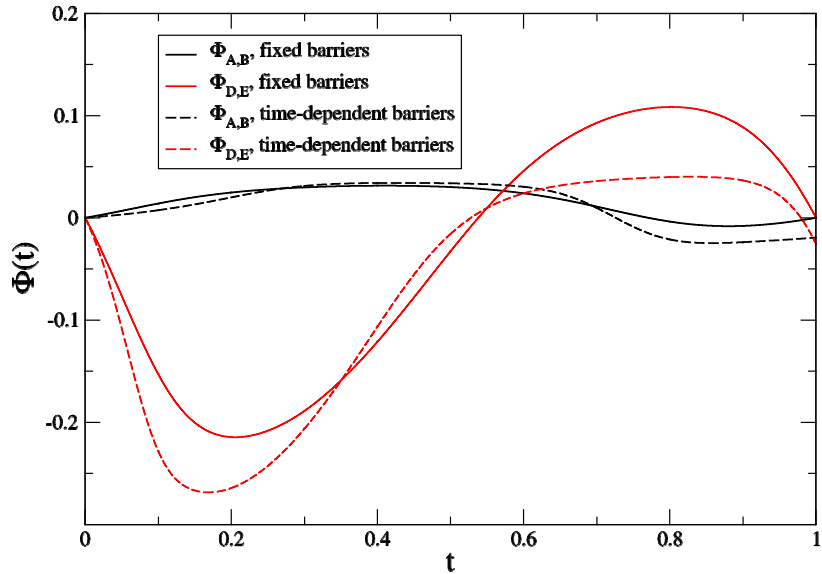}

\caption{Comparison of time-integrated currents with time-independent (solid lines) and time dependent barriers (dashed lines), for an
interaction with $U_{1/2}=U_{3/2}=U_{5/2}=0$. The results depicted in the figure show that the no-pumping theorem holds, namely that $\Phi_{X,Y}(1)=0$ for fixed barriers, as predicted by
the theory presented in Sec. \ref{sec:anal} \label{fig:barriers}}
\end{figure}

To check the validity of the no-pumping theorem the integrated currents $\Phi_{A,B} (t)\equiv \int_{0}^{t} dt^\prime J^{ps}_{B\rightarrow A} (t^\prime)$ and $\Phi_{D,E} (t)$
were calculated numerically. Choosing $U_{1/2}=U_{3/2}=U_{5/2}=0$, with $U_0=1$, $U_1=-0.3$, and $U_2=-0.6$,
allows to examine a finite-range interaction of the type discussed in Sec. \ref{sec:anal}, where it was predicted to satisfy the
no-pumping theorem.
The two-particle pump was driven by a cyclic variation of site energies using $E_A(t)=\cos (2 \pi t)$, $E_B(t)=2+1.1 \cos (2 \pi t +\pi/4)$,
$E_C(t)=1+\cos(2 \pi t+\pi/2)/2$, $E_D(t)=1/2+\cos(2 \pi t +\pi)$, and $E_E (t)= 2+2 \cos (2 \pi t +3 \pi/2)$. This prediction is
tested in Fig. \ref{fig:barriers}, which compares the integrated particle fluxes for fixed and time-dependent local barriers.
The solid lines correspond to a stochastic pump with time-independent local barriers. The values $B_{A,B}=1.5$, $B_{B,C}=0.7$, $B_{A,C}=0.3$, $B_{A,E}=0.25$,
$B_{C,D}=1.7$, and $B_{D,E}=0.1$ were used. The dashed lines depict the integrated currents for a system with time-dependent barriers. Here three
of the barriers were varied according to $B_{A,B}(t)= 1.5+\cos(4 \pi t)$, $B_{D,E}(t)=2 \cos(2 \pi t +\pi/2)$, and $B_{C,D}=1/2+\cos(2 \pi t)/2$, while
the other barriers retained their time-independent value.

The no-pumping theorem concerns with the net accumulation of particles after a full cycle. One should therefore examine the value of $\Phi(t=1)$ to see if it vanishes or not. The results depicted
in Fig. \ref{fig:barriers} show that after a full cycle the particle fluxes vanish for the case of time-independent local barriers. In contrast, there is a net particle flux when the local barriers are varied in
time. This matches the prediction of the no-pumping theorem. One should keep in mind that the interaction between particles used to generate Fig. \ref{fig:barriers} in non-local.
The results include the time-integrated particle fluxes for intermediate times and not only after a full cycle to demonstrate that these fluxes vary with time in a non-trivial manner, and vanish only after a full cycle.

\begin{figure}[h]
\includegraphics[scale=0.45]{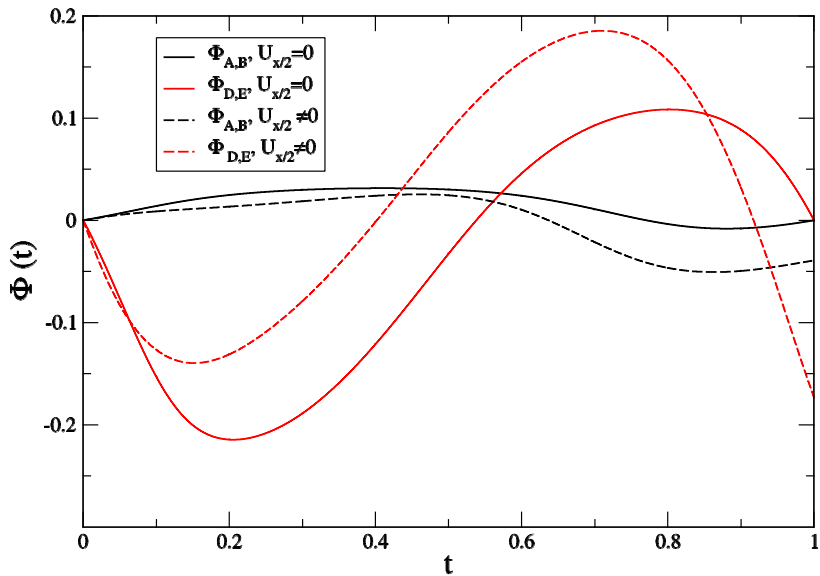}

\caption{Testing the validity of the no-pumping theorem with interaction dependent barriers. Solid lines correspond to many-particle barriers that do not depend on the state of the system, that is to the case of $U_{1/2}=U_{3/2}=U_{5/2}=0$. Dashed lines correspond to many-particle barriers that are affected by the interaction as described in the text. All the curves were
calculated for time-independent $B_{X,Y}$. The system with the interaction-dependent barriers clearly violates the no-pumping theorem. \label{fig:uhalf}}
\end{figure}

To test the possible influence of an interaction that affect the jumping particle I compared the dynamics of a pump without barrier interactions,
$U_{1/2}=U_{3/2}=U_{5/2}=0$, to that of a pump with interaction dependent barriers. The latter were chosen as  $U_{1/2}=-2$, $U_{3/2}=0$, and $U_{5/2}=3$. Here the barriers
 $B_{X,Y}$ were taken to be time-independent, since the no-pumping theorem tells us that directed motion will appear for time-dependent barriers independently of the interaction. Beside the barrier interactions all other parameters, namely the local barriers, interactions, and site energies used in the numerics are exactly those
used in the fixed barrier case presented in Fig. \ref{fig:barriers}.
The results of the comparison are depicted in Fig. \ref{fig:uhalf}.

One immediately notices that the inclusion of barrier interactions leads to the breakdown of the no-pumping theorem. Such interactions were not considered
in the derivation of Sec. \ref{sec:anal}, so there is no contradiction with the conclusions obtained there. Instead, one should view this numerical result as evidence that
it is precisely the presence of interactions in the many-particle barriers
results in violation of the no-pumping theorem.

\section{Discussion}
\label{sec:disc}

The results presented in Secs. \ref{sec:anal} and \ref{sec:num} help to clarify the relationship between the properties
of particle-particle interactions and the validity of the no-pumping theorem. Specifically, the results indicate that
no-pumping holds for quite general interactions, as long as they affect only the many-particle energies. Once the many-body interaction
enters into the transition barriers in a non trivial way, as happened for the model with $U_{j/2}\ne 0$ that was studied in Sec. \ref{sec:num},
no-pumping is likely to be violated. While this gives an answer to the question posed in the introduction, it is worth to
take a closer look at the mechanism through which these interactions break the no-pumping theorem.

A better understanding of the reason for violation of no-pumping can be gained by trying to
reproduce the proof presented in Sec. \ref{sec:anal} using the model studied in Sec. \ref{sec:num} and seeing where
the derivation breaks down. Since particle conservation holds for any interaction, the problem must arise
in the derivation of cycle equations. Let us consider all the possible cycles of transitions between many-particle
states in which one particle makes the series of transitions $A\rightarrow B \rightarrow C \rightarrow A$ while the other
is a stationary spectator.
Without affecting the generality of the argument, let us assume that the first particle is the spectator. When it is located at site $A$, one finds
\begin{equation*}
  e^{\beta B_{A,B}+\beta U_{1/2}} J_{A,A\rightarrow B} (t) + e^{\beta B_{B,C}+\beta U_{3/2}} J_{A, B\rightarrow C} (t) + e^{\beta B_{A,C}+\beta U_{1/2}} J_{A, C\rightarrow A} (t) = 0.
\end{equation*}
When the spectator is located at site $B$ one finds
\begin{equation*}
   e^{\beta B_{A,B}+\beta U_{1/2}} J_{B,A\rightarrow B} (t) + e^{\beta B_{B,C}+\beta U_{1/2}} J_{B,B\rightarrow C} (t) + e^{\beta B_{A,C}+\beta U_{3/2}} J_{B,C\rightarrow A} (t) = 0.
\end{equation*}
One can derive similar equations for cycles in which the spectator is at $C$, $D$, or $E$.

The derivation of the cycle equations (\ref{eq:cycles}) involved a summation of equations of this type. A crucial part of the derivation was
the identification of certain sums of currents $J_{X,Y\rightarrow Z}$ as the total particle current between sites. Such an identification was possible
because all contributing currents came with the same prefactor, see Eq. (\ref{eq:totcurrent}) or Eq. (\ref{eq:totcurrent2}). One immediately sees
that the $U_{j/2}$ part of the interaction gives different prefactors to currents in the expressions above, and therefore
prevents one from combining these equation in a way that would result in cycle equations for total particle currents. As a result there are no simultaneous
system of cycle equations and conservation laws for the same type of fluxes - the total particle current in this setup - and no-pumping is not ensured. The numerical results presented in Sec. \ref{sec:num}
explicitly show that no-pumping is violated in such cases.

One way of interpreting the way that interaction affects the dynamics of a stochastic pumps is by examining the so-called
branching fractions. The branching fractions are defined as the conditional probability that a particle will make
a specific transition, e.g. $X \rightarrow Y$, when it is known that the particle is making a transition out of site $X$. Branching fractions therefore express the probability
of ending in different sites in a transition.

Consider the model studied in Sec. \ref{sec:num}. Let us calculate the probability that a particle ends at a site $B$, given that
it leaves site $C$. This branching fraction will depend on the location of the other particle (namely, the spectator).
A short calculation shows that this branching fraction is
\begin{equation*}
  P(C\rightarrow B|C)=\frac{e^{-\beta B_{B,C}}}{e^{-\beta B_{A,C}}+e^{-\beta B_{B,C}}+e^{-\beta B_{C,D}}},
\end{equation*}
when the spectator is at $C$. In contrast, when the spectator is at $A$ the branching fraction
is
\begin{equation*}
  P(C\rightarrow B|A)=\frac{e^{-\beta B_{B,C}}}{e^{-\beta \left(B_{A,C}+U_{1/2}-U_{3/2} \right)}+e^{-\beta B_{B,C}}+e^{-\beta B_{C,D}}}.
\end{equation*}
This demonstrates that interactions like the one studied in Sec. \ref{sec:num} result in branching fractions that depend on the many-particle state
of the system, expressed here through the dependence on the location of the spectator.
This gives the branching fractions an effective time dependence, which is mediated through the evolution
of the many-particle probability distribution during the driving cycle. At different times during the cycle different
many-particle states are more likely, and with them different values of branching fractions.

A point worth emphasizing is that for single-particle stochastic pumps, as well as for pumps of particles with zero-range interactions, the
condition on time independence of barriers and of branching fractions are essentially equivalent. (Strictly speaking one can
add a time-dependent term to all barriers without affecting branching fractions.)
This means that previous work could have used branching fractions
to express the conditions for no-pumping. Indeed,
Maes {\it et al.} have presented a derivation
of the no-pumping theorem for non-Markovian systems which uses conditions on the time dependence of branching fractions \cite{maes_general_2010}.
The condition that the branching fractions will be time independent is equivalent to the so called
``direction-time independence'' that was used in other contexts in semi-Markov processes \cite{Qian2006,Andrieux2008}.

The results presented in Sec. \ref{sec:anal} demonstrate that in models where the interaction does not affect the branching fractions
the no-pumping theorem holds. In contrast, the example studied in Sec. \ref{sec:num} shows that no-pumping breaks down once the branching
fractions depend on the state of the system. While this is a numerical result for a single model, and not a general proof, its connection to
the known role that branching fractions play in no-pumping strongly suggests that the results are quite general. Namely, it suggests that a typical
interaction that results in state-dependent branching fractions would lead to the breakdown of the no-pumping theorem.

Let us briefly return to the paradigmatic examples of zero-range and exclusion interactions discussed earlier. A stochastic pump
of particles with zero-range interactions will have state-independent branching fractions, and indeed it is known that such systems
satisfy the no-pumping theorem \cite{asban_no-pumping_2014}. In contrast, systems with exclusion interactions between particles must have
state-dependent branching fractions, simply because the particles block each other. The results presented here mean that it is precisely this
property of the interaction, and not its finite range, that causes no-pumping to be violated.

\begin{acknowledgments}
I would like to thank Ori Hirschberg for illuminating discussions that have
initiated my interest in this topic.
This work was supported by the the U.S.-Israel Binational Science
Foundation (Grant No. 2014405), by the Israel Science Foundation (Grant
No. 1526/15), and by the Henri Gutwirth Fund for the Promotion of
Research at the Technion.
\end{acknowledgments}

\end{document}